\documentclass[aps,floats]{revtex4}
\usepackage{amsmath,amssymb}
\usepackage{graphicx,epsfig}

\begin{document}
\bibliographystyle {plain}

\def\oppropto{\mathop{\propto}} 
\def\opsimeq{\mathop{\simeq}}
\def\opoverderline{\mathop{\overline}}
\def\operarrow{\mathop{\longrightarrow}}
\def\opsim{\mathop{\sim}}

\def\fig#1#2{\includegraphics[height=#1]{#2}}
\def\figx#1#2{\includegraphics[width=#1]{#2}}


\title{ Dynamical barriers for the random ferromagnetic Ising model on the Cayley tree :
 \\ traveling-wave solution of the real space renormalization flow } 


 \author{ C\'ecile Monthus and Thomas Garel }
  \affiliation{ Institut de Physique Th\'{e}orique, CNRS and CEA Saclay,
 91191 Gif-sur-Yvette, France}

\begin{abstract}

We consider the stochastic dynamics near zero-temperature of the random ferromagnetic Ising model on a Cayley tree of branching ratio $K$. We apply the Boundary Real Space Renormalization procedure introduced in our previous work (C. Monthus and T. Garel, J. Stat. Mech. P02037 (2013)) in order to derive the renormalization rule for dynamical barriers. We obtain that the probability distribution $P_n(B)$ of dynamical barrier for a subtree of $n$ generations converges for large $n$ towards some traveling-wave $P_n(B) \simeq P^*(B-nv) $, i.e. the width of the probability distribution remains finite around an average-value that grows linearly with the number $n$ of generations. We present numerical results for the branching ratios $K=2$ and $K=3$. We also compute the weak-disorder expansion of the velocity $v$ for $K=2$.

\end{abstract}

\maketitle

\section{ Introduction }

The stochastic dynamics of classical disordered spin systems turns out to be extremely slow in the whole low-temperature phase (see for instance the books \cite{young,houches} and references therein). 
The reason is that the system tends to remain trapped in valleys of configurations on various scales.
Within the point of view of dynamical simulations, this problem has been called the 
'futility' problem \cite{werner} : 
the number of distinct configurations visited during the simulation remains very small 
with respect to the accepted moves. The reason is that the system
 visits over and over again the same configurations
 within a given valley before it is able to escape towards another valley.

As a consequence, a natural idea is
to formulate some appropriate renormalization procedure for dynamical barriers.
For random walks in random media, Strong Disorder Renormalization rules have been formulated
in {\it real space } \cite{sinairg,sinaibiasdirectedtraprg,us_2d}. 
The direct generalization of this approach to many-body systems leads to Strong Disorder Renormalization formulated in {\it configuration space } \cite{us_rgconfig}, because the renormalization concerns the master equation of the dynamics defined in configuration space. 
These Strong Disorder RG approaches are perfect to describe correctly the hierarcal organization of valleys within valleys.
However, from a numerical point of view, since the size of the configuration space grows exponentially
 with the number of degrees of freedom of the many-body system, this approach can be applied numerically only for small sizes \cite{us_rgconfig}.

In a recent work \cite{us_rgdyn}, we have thus introduced a different renormalization approach
formulated in {\it real space } : using the standard mapping between 
the detailed-balance dynamics of classical Ising models and some quantum Hamiltonian,
we have derived appropriate real-space renormalization rule for this quantum Hamiltonian.
We have solved explicitly the renormalization flow for the random ferromagnetic chain 
 \cite{us_rgdyn}, for the pure Ising model on the Cayley tree  \cite{us_rgdyn}, and for
the hierarchical Dyson Ising model \cite{us_rgdyndyson}. In the present paper, we study
the case of the random ferromagnetic Ising model on the Cayley tree.

The paper is organized as follows.
In section \ref{sec_rg}, we explain how the Real Space Renormalization approach of \cite{us_rgdyn}
can be applied to the stochastic dynamics of the
random ferromagnetic Ising model on the Cayley tree.
In section \ref{sec_rgbarrier}, we derive the renormalization rules for dynamical barriers near zero temperature. In section \ref{sec_rgflow}, we study the renormalization flow for the probability distribution of dynamical barriers.
The cases of branching ratio $K=2$ and $K=3$ are discussed respectively in sections
\ref{sec_k2} and \ref{sec_k3} with numerical results on the traveling-wave statistics of dynamical barriers. Section \ref{sec_conclusion} summarizes our conclusions.

\section{  Real Space Renormalization Approach }

\label{sec_rg}

\subsection{ Model and notations }

We consider the random ferromagnetic Ising model with the classical energy
\begin{eqnarray}
U({\cal C}) = -\sum_{i<j} J_{ij} S_i S_j
\label{Uspin}
\end{eqnarray}
 defined on a Cayley tree of branching ratio $K$ with $N$ generations, and
with free boundary conditions on all the boundary spins. The coordinence of non-boundary spins is thus $(K+1)$.
The couplings $J_{ij}$ are independent random positive variables drawn with some law $\rho(J)$.

The stochastic dynamics is defined by the master equation 
\begin{eqnarray}
\frac{ dP_t \left({\cal C} \right) }{dt}
= \sum_{\cal C '} P_t \left({\cal C}' \right) 
W \left({\cal C}' \to  {\cal C}  \right) 
 -  P_t \left({\cal C} \right) W_{out} \left( {\cal C} \right)
\label{master}
\end{eqnarray}
that describes the time evolution of the
probability $P_t ({\cal C} ) $ to be in  configuration ${\cal C}$
 at time t.
The notation $ W \left({\cal C}' \to  {\cal C}  \right) $ 
represents the transition rate per unit time from configuration 
${\cal C}'$ to ${\cal C}$, and 
\begin{eqnarray}
W_{out} \left( {\cal C} \right)  \equiv
 \sum_{ {\cal C} '} W \left({\cal C} \to  {\cal C}' \right) 
\label{wcout}
\end{eqnarray}
represents the total exit rate out of configuration ${\cal C}$.
We will focus here on single spin-flip dynamics satisfying
detailed balance
\begin{eqnarray}
e^{- \beta U({\cal C})}   W \left( \cal C \to \cal C '  \right)
= e^{- \beta U({\cal C '})}   W \left( \cal C' \to \cal C   \right)
\label{detailed}
\end{eqnarray}
 where the transition rate corresponding to the flip of a single spin $S_k$ reads
\begin{eqnarray}
W \left( S_k \to -S_k \right)
=G^{ini} \left[ h_k=\sum_{i \ne k} J_{ik} S_i \right]   e^{ - \beta S_k \left[ \sum_{i \ne k} J_{ik} S_i \right] }
\label{WG}
\end{eqnarray}
$G^{ini}[h_k]$ is an arbitrary positive even function of the local field $h_k=\sum_{i \ne k} J_{ik} S_i $.
For instance, the Glauber dynamics corresponds to the choice
\begin{eqnarray}
G^{ini}_{Glauber} [h] = \frac{1}
{ 2 \cosh \left( \beta  h \right) }
\label{glauber}
\end{eqnarray}

\subsection{ Associated quantum Hamiltonian }

As is well-known, the master equation of Eq. \ref{master} with the choice of Eq. \ref{WG}
can be mapped via the similarity transformation $
P_t ( {\cal C} ) \equiv e^{-  \frac{\beta}{2} U(\cal C ) } \psi_t ({\cal C} ) $ onto a Schr\"odinger 
equation with the following quantum Hamiltonian involving Pauli matrices $(\sigma^x,\sigma^z)$
\cite{glauber,felderhof,siggia,kimball,peschel,us_conjugate,castelnovo,us_rgdyn}
\begin{eqnarray}
{\cal H } && 
=  \sum_{ k } G^{ini} \left[ \sum_{i \ne k} J_{ik} \sigma^z_i \right]
  \left( e^{ - \beta \sigma^z_k \left( \sum_{i \ne k} J_{ik} \sigma^z_i \right) }
-   \sigma^x_k \right)
\label{Hgene}
\end{eqnarray}

Note that in the high-temperature limit $\beta \to 0 $,
the quantum Hamiltonian of Eq. \ref{Hgene} for the Glauber dynamics of Eq. \ref{glauber}
reduces to the standard transverse-field Ising model 
\begin{eqnarray}
{\cal H }
\opsimeq_{\beta \to 0 } N + \sum_{ k=1 }^N   \left(  - \beta \sigma^z_k \left( \sum_{i \ne k} J_{ik} \sigma^z_i \right) 
-   \sigma^x_k \right)
\label{HhighT}
\end{eqnarray}
When the coupling $J_{ij}$ are random, the low-energy physics is then well described
by the Strong Disorder RG procedure valid both in one dimension \cite{danieltransverse}
and in higher dimensions $d>1$ \cite{fisherreview,motrunich,kovacsreview} (see \cite{StrongRGreview} for a review).
However here we are interested into the opposite limit of very low temperature 
where $\beta \to +\infty $, where one cannot linearize the exponentials in the quantum Hamiltonians of Eq. \ref{Hgene}.
We have explained in \cite{us_rgdyn} how to
 define appropriate real-space renormalization
rules for this type of quantum Hamiltonian in the opposite limit of very low temperature.
In the following, we recall the main idea for the case of the Cayley tree geometry.

\subsection{ Reminder on the Boundary Real Space Renormalization on a Cayley tree } 

\label{sec_reminder}

\begin{figure}[htbp]
 \includegraphics[height=8cm]{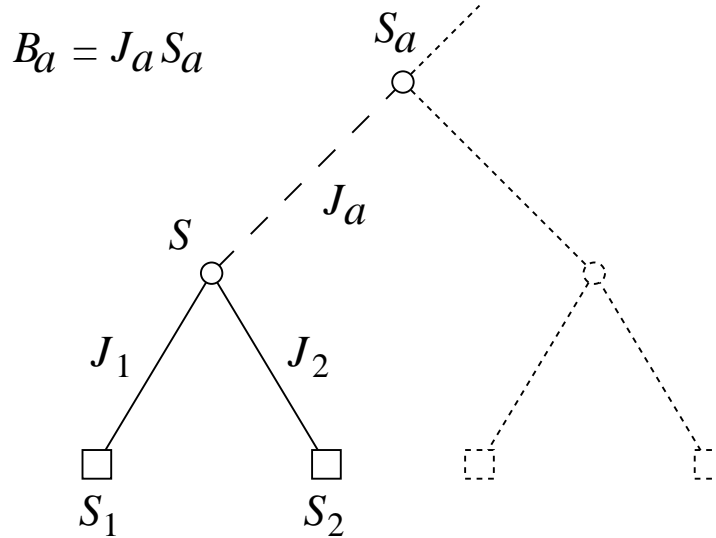}
\caption{ Basic step of the Boundary Real Space Renormalization for a Cayley tree of branching ratio $K=2$: the quantum Hamiltonian of Eq. \ref{HK} involves $K+1=3$ spins, namely the two renormalized boundary spins $(S_1,S_2)$ (that represent sub-trees) whose renormalized dynamics is described by renormalized amplitudes $(G_1,G_2)$, and their common ancestor $S$, whose dynamics is still described by the initial amplitude $G^{ini}[h]$. The further ancestor $S_a$ is not taken into account as a quantum spin, but only through
the external field $B_a=J_a S_a$ seen by the spin $S$.
}
\label{figboundaryrg}
\end{figure}

For a model defined on a Cayley tree, it is natural to define a 
{\it boundary renormalization } procedure in order to keep the tree topology unchanged,
and to obtain recurrence equations.
For instance for the Random Transverse Field Ising Model of Eq. \ref{HhighT},
this idea has been used either within the Quantum Cavity Approach  \cite{ioffe,feigelman,dimitrova}
or within the Boundary RG approach  \cite{us_transtree}.

For the Hamiltonian of Eq. \ref{Hgene},
 a boundary spin $S_i$ is connected to a single ancestor spin $S$ via some 
random positive coupling $J_i>0$,
so the absolute value of its local field $h_i=J_i S$ takes the single value $J_i$.
As a consequence, the function $G^{ini}(h_i)$ reduces to the number $G^{ini}(J_i)$.
As explained in \cite{us_rgdyn} for the case of the pure Ising model,
 it is thus convenient to define a Boundary Real Space Renormalization as follows.
The basic renormalization step concerns 
$K$ renormalized boundary spins $(S_1,S_2,..,S_K)$
whose renormalized dynamics is described by some renormalized amplitudes $G_i$ 
(which are numbers and not operators) and their common ancestor spin $S$
whose dynamics is still described by the initial amplitude $G^{ini}[h]$ involving also
the external field $B_a=J_a S_a$ induced by its next ancestor spin $S_a$.
So we have to study the following effective
Hamiltonian for these $(K+1)$ spins $(S_1,..,S_K,S)$
\begin{eqnarray}
 H_{K+1} && = G^{ini} \left[ \sum_{i=1}^K  J_{i} \sigma^z_{i} +B_a \right]
 \left( e^{ - \beta  \sigma^z (\sum_{i=1}^K  J_{i} \sigma^z_{i} +B_a )  } -   \sigma^x \right)
+   \sum_{i=1}^K G_{i} 
 \left( e^{ - \beta \sigma^z_{i}  J_{i} \sigma^z)  } -   \sigma^x_{i} \right)
\label{HK}
\end{eqnarray}
The physical meaning of the amplitude $G_i$ of the renormalized boundary spin $S_i$
that represents a whole sub-tree, 
is that the largest relaxation time for this isolated sub-tree reads
(see detailed explanations in \cite{us_rgdyn})
\begin{eqnarray}
t^{relax}_i = \frac{1}{ 2 G_i} 
\label{taueqGfinal}
\end{eqnarray}
Near zero temperature, the largest relaxation time is simply 
the equilibrium time $t^{eq}_i$ needed to flip between the two renormalized states
states $S_i=+1$ and $S_i=-1$ representing the two ferromagnetic ground states of the 
corresponding sub-tree, so the corresponding dynamical barrier ${\cal B}_i $ 
 is defined
by the exponential behavior
\begin{eqnarray}
t^{relax}_i = \frac{1}{ 2 G_i} \oppropto_{ \beta \to +\infty} e^{ \beta {\cal B}_i }
\label{bi}
\end{eqnarray}

After the first renormalization step, the renormalized dynamical barrier ${\cal B}_i $
are expected to grow, so that
the renormalized amplitudes $G_i$ will be extremely small.
The most appropriate approach is then a perturbative analysis
in the parameters $G_i$ that may be summarized as follows (see \cite{us_rgdyn}
for more details) :

(i) When $G_i=0$ for $i=1,2,..,K$, the spins $(S_1,..,S_K)$ cannot flip and are thus frozen.
So the $2^K$ states 
\begin{eqnarray}
\vert v^{S_1,S_2,..,S_K}_0 > && \equiv \prod_{j=1}^K \vert S_j>  \sum_{S=\pm} 
\frac{e^{ \frac{\beta}{2} S (\sum_{i=1}^K J_i S_i+B_a) }}{ \sqrt{ 2 \cosh \beta (\sum_{i=1}^K J_i S_i+B_a)} }  \vert S > 
\label{vkbtree}
\end{eqnarray}
are zero-energy states of Hamiltonian of Eq. \ref{HK} when $G_i=0$
for any function $G^{ini}[h]$ since one has
\begin{eqnarray}
 \left( e^{ - \beta \sigma^z \left( \sum_{i=1}^K J_i \sigma^z_i+B_a \right) }-   \sigma^x \right)
\vert v^{S_1,S_2,..,S_K}_0 > && =0
\label{HKtreev}
\end{eqnarray}
The physical interpretation is that the spin $S$ is at equilibrium with respect to the
frozen spins $(S_1,..,S_K)$.
The other $2^K$ states have a finite energy for $G_i=0$.

(ii) When the amplitudes $G_i$ for $i=1,2,..,K$ are small, 
 we need to diagonalize the perturbation within the subspace 
spanned by the $2^K$ vectors $\vert v^{S_1,S_2,...S_K}_0 >$ of Eq. \ref{vkbtree}.
 We look for an eigenstate via the linear combination
\begin{eqnarray}
\vert u_{\lambda} > && = \sum_{S_1=\pm,S_2=\pm,..,S_K=\pm} T_{\lambda}(S_1, S_2,...,S_K) \vert v^{S_1,S_2,...S_K}_0 >
\label{vktree}
\end{eqnarray}
The eigenvalue equation $0=(H_{K+1}-\lambda) \vert u_{\lambda} >$ reads
\begin{eqnarray}
0  && =  
 \left[  \sum_{i=1}^K G_{i} \frac{ 2 \cosh \beta  (\sum_{j\ne i} J_jS_j+B_a) }{  2 \cosh \beta (\sum_{j=1}^K J_j S_j+B_a) } -\lambda_1  \right] T_{\lambda_1}(S_1,,...,S_K)
\nonumber \\
&& -  \sum_{i=1}^K G_{i} \frac{ 2 \cosh  \beta   (\sum_{j\ne i} J_jS_j+B_a) }
{\sqrt{ 2 \cosh \beta (\sum_{j\ne i} J_jS_j -J_i S_i
+B_a)}  \sqrt{ 2 \cosh \beta (\sum_{j\ne i} J_j S_j+J_iS_i+B_a)} }
T_{\lambda_1}(S_1, .. ,-S_i,...,S_K)
\label{eigenperturbprojexpli}
\end{eqnarray}

(iii) As explained in detail in \cite{us_rgdyn}, one obtains 
that near zero-temperature, the Hamiltonian of Eq. \ref{HK} can be renormalized onto
the single spin effective Hamiltonian
\begin{eqnarray}
H \simeq G_R  \left( e^{ - \beta B_a \sigma^z_R  }
-   \sigma^x_R \right)
\label{Hefflowestfinal}
\end{eqnarray}
for the renormalized spin (describing the full ferromagnetic states)
\begin{eqnarray}
\vert S_R=+> && \equiv  \left(  \prod_{j=1}^K \vert S_j=+> \right)   \vert S= + > \nonumber \\
\vert S_R=-> && \equiv  \left(  \prod_{j=1}^K \vert S_j=-> \right)   \vert S= - > 
\label{SR}
\end{eqnarray}
with the renormalized amplitude (using Eq \ref{lambda1final})
 \begin{eqnarray}
G_R &&  =\frac{ \lambda_1 }
{ 2 \cosh    \beta  J_a   } 
\label{ggenetree}
\end{eqnarray}
where $\lambda_1$ is the first non-vanishing eigenvalue of the system
of Eq. \ref{eigenperturbprojexpli}.

In the pure case studied in \cite{us_rgdyn}, i.e. when
 all couplings $J_i$ take the same value $J$,
and where all amplitudes $G_i$ take the same value $G$,
we have solved Eq \ref{eigenperturbprojexpli} by taking into account
the symmetry between the $K$ branches.
 Here in the disordered case, the $K$ branches are not equivalent anymore,
but we can nevertheless derive an explicit renormalization rule for $G_R$,
as explained in the following section.

\section { Renormalization rule for dynamical barriers }

\label{sec_rgbarrier}

\subsection{ Eigensystem for $\lambda_1$  }

To see more clearly the meaning of the eigensystem of Eq. \ref{eigenperturbprojexpli},
it is convenient to introduce  
\begin{eqnarray}
T_0(S_{1},..,S_{K}) \equiv  \sqrt{2 \cosh \beta (\sum_{i=1}^K  J_{i} S_{i}+B_a )  }  
\label{psizeroKtreevcompob}
\end{eqnarray}
which solve the system of Eq. \ref{eigenperturbprojexpli} for $\lambda=0$,
and the ratios
\begin{eqnarray}
A(S_{1},..,S_{K})  \equiv 
\frac{ T_{\lambda_1}(S_{1},..,S_{K})}
{ T_{0}(S_{1},..,S_{K}) }   
\label{defamplirA}
\end{eqnarray}
so that Eq \ref{eigenperturbprojexpli} reads
\begin{eqnarray}
&& 0    = 
\left[  \sum_{i=1}^K G _{i}  \frac{T_0^2(S_{1},.,S_i=0, .,S_{K})  }
{ T_0^2(S_{1},.,S_i, .,S_{K})  } -\lambda  \right]  A(S_1,,...,S_K)
 -  \sum_{i=1}^K G_{i} \frac{ T_0^2(S_{1},.,S_i=0, .,S_{K}) }
{  T_0^2(S_{1},.,S_i,.,S_{K})    } 
 A(S_{1},..,-S_{i},..,S_{K}) 
\label{eigenperturbprb}
\end{eqnarray}

To compute the lowest non-vanishing eigenvalue $\lambda_1$ of Eq. \ref{eigenperturbprojexpli},
it is consistent to set $\lambda_1=0$ in all equations 
except at the two extreme cases  where all spins have the same value, either $S_a$ or $-S_a$,
where $B_a\equiv J_a S_a$.
The two extreme components should be orthogonal to Eq \ref{psizeroKtreevcompob}
and thus read
\begin{eqnarray}
T_{\lambda_1}(S_a,S_a,...,S_a) && = T_0 (-S_a,-S_a,...,-S_a)    = 
\sqrt{2 \cosh \beta (\sum_{i=1}^K  J_{i}  -J_a  )  } 
\simeq e^{ \frac{\beta}{2} \left( \sum_{i=1}^K  J_{i}-J_a\right)  }
\nonumber \\
T_{\lambda_1}(-S_a,-S_a,...,-S_a)  && = -T_{0}(S_a,S_a,...,S_a)    = - 
  \sqrt{2 \cosh \beta (\sum_{i=1}^K  J_{i}+J_a ) }
 \simeq - e^{ \frac{\beta}{2} \left( \sum_{i=1}^K  J_{i}+J_a\right)  }
\label{psiun3treecompob}
\end{eqnarray}
i.e. the two extreme values of the ratios of Eq. \ref{defamplirA} 
\begin{eqnarray}
A(S_a,S_a,...,S_a) && = \frac{T_0(-S_a,-S_a,...,-S_a)}{T_{0}(S_a,S_a,...,S_a)}    =
 \frac{\sqrt{2 \cosh \beta (\sum_{i=1}^K  J_{i}-J_a  )  }}{\sqrt{2 \cosh \beta (\sum_{i=1}^K  J_{i}+J_a  ) }} \simeq e^{- \beta J_a}
\nonumber \\
A(-S_a,-S_a,...,-S_a) && = - \frac{ T_{0}(S_a,S_a,...,S_a) }{T_0(-S_a,-S_a,...,-S_a)}   = - 
\frac{\sqrt{2 \cosh \beta (\sum_{i=1}^K  J_{i}+J_a  ) }}{\sqrt{2 \cosh \beta (\sum_{i=1}^K  J_{i}-J_a  )  }}
\simeq - e^{ \beta J_a}
\label{aextremesb}
\end{eqnarray}
satisfy Eq \ref{eigenperturbprb} with $\lambda_1$. So 
$\lambda_1$ can be computed near zero temperature as
\begin{eqnarray}
  \lambda_1  &&   =
  \sum_{i=1}^K G _{i} \frac{T_0^2(S_a,S_a,...,S_a,S_i=0, S_a,S_a,...,S_a)  }
{ T_0^2(S_a,S_a,...,S_a)  }  \left(1- \frac{ A(S_a,S_a,...,S_a,S_{i}=-S_a,S_a,S_a,...,S_a) }{ A(S_a,S_a,...,S_a)}  \right)
\nonumber \\
 &&   \simeq
  \sum_{i=1}^K G _{i} e^{- \beta J_i}
  \left(1- \frac{ A(S_a,S_a,...,S_a,S_{i}=-S_a,S_a,S_a,...,S_a) }{ A(S_a,S_a,...,S_a)}  \right)
\label{eigenperturbprprojexpliextremb}
\end{eqnarray}
whereas 
all non-extreme values satisfy Eq \ref{eigenperturbprb} with $\lambda_1=0$, 
so we may drop the common denominator $T_0^2(S_{1},.,.,S_{K})   $ to obtain
\begin{eqnarray}
 0  = \left[  \sum_{i=1}^K G _{i}  T_0^2(S_{1},.,S_i=0, .,S_{K})    \right]  A(S_1,,...,S_K)
 -  \sum_{i=1}^K G_{i}   T_0^2(S_{1},.,S_i=0, .,S_{K}) 
 A(S_{1},..,-S_{i},..,S_{K})  
\label{eigenperturbprbulkb}
\end{eqnarray}
In the disordered case, we expect that the dynamical transition between the two 
ferromagnetic states will be dominated by a single dynamical path near zero temperature.
So let us now compute the dynamical barrier associated
 to a given dynamical path.

\subsection{ Dynamical barrier associated to a given dynamical path  $(1,2,..,K)$}

In this section, we consider the given dynamical path called $(1,2,..,K)$
\begin{eqnarray}
A(-S_a,-S_a,-S_a,...,-S_a,-S_a) \leftrightarrow A(S_a,-S_a,-S_a,-S_a,-S_a) \leftrightarrow A(S_a,S_a,-S_a,...,-S_a,-S_a)  \nonumber \\
...   \leftrightarrow A(S_a,S_a,...,S_a,-S_a) \leftrightarrow A(S_a,S_a,S_a,S_a,S_a)
\end{eqnarray}
Then we are left with a one-dimensional problem with the notations 
\begin{eqnarray}
q(0) && = A(-S_a,-S_a,-S_a...,-S_a)  \simeq - e^{ \beta J_a}
 \nonumber \\
q(1) && = A(S_a,-S_a,-S_a,...,-S_a)
 \nonumber \\
q(2) && = A(S_a,S_a,-S_a,...,-S_a)
 \nonumber \\
...
\nonumber \\
q(K-1) && = A(S_a,S_a,...,S_a,-S_a) 
 \nonumber \\
q(K) && =A(S_a,S_a,... S_a,S_a) \simeq e^{- \beta J_a}
\label{smallq}
\end{eqnarray}
 Eq \ref{eigenperturbprprojexpliextremb} reduces to
\begin{eqnarray}
 \lambda_1  &&   
  \simeq  G _{K} e^{- \beta J_K}  \left(1- \frac{ q(K-1) }{ q(K) }  \right)
\label{eigenperturbprojexpliextremfinal}
\end{eqnarray}
and Eq. \ref{eigenperturbprbulkb} becomes for $1 \leq k \leq K-1$
\begin{eqnarray}
  q(k)   = p_-(k)  q(k-1) +p_+(k) q(k+1) 
\end{eqnarray}
with the probabilities 
\begin{eqnarray}
p_-(k) && = \frac{G_{k}   T_0^2(S_a,...,S_a,S_k=0,-S_a,-S_a,...,-S_a)}
{G_{k}   T_0^2(S_a,...,S_a,S_k=0,-S_a,...,-S_a) 
 + G_{k+1}  T_0^2(S_a,,...,S_a,S_{k+1}=0,-S_a,...,-S_a) }
 \nonumber \\
p_+(k) && = \frac{ G_{k+1}   T_0^2(S_a,...,S_a,S_{k+1}=0,-S_a,,...,-S_a)}
{G_{k}   T_0^2(S_a,,...,S_a,S_k=0,-S_a,...,-S_a) 
 + G_{k+1}   T_0^2(S_a,,...,S_a,S_{k+1}=0,-S_a,...,-S_a) }
\label{ppm}
\end{eqnarray}
normalized to $p_-(k)+p_+(k)=1$.

The solution can be obtained by \cite{rec1d,us_rgdyn}
\begin{eqnarray}
 q(k)= q(0)\frac{R_0(k,K-1)}{R_0(0,K-1)}  +q(K)  \frac{R_K(1,k)}{R_K(1,K)}
\label{soluqk}
\end{eqnarray}
using Kesten variables \cite{kestenv}
\begin{eqnarray}
R_0(K,K-1) && =0  \\
R_0(K-1,K-1) && =1 \nonumber \\
R_0(k \leq K-2 , K-1) && =1+\sum_{m=k+1}^{K-1} \prod_{n=m}^{K-1} \frac{ p_+(n)}{ p_-(n)}
\nonumber \\
R_0(0,K-1) && = 1+\sum_{m=1}^{K-1} \prod_{n=m}^{K-1} \frac{ p_+(n)}{ p_-(n)}
 = 1+ \frac{p_+(K-1)}{p_-(K-1)} 
+ ... +
 \frac{p_+(K-1) p_+(K-2)...p_+(1)}{p_-(K-1) p_-(K-2)...p_-(1)}
\nonumber 
\label{rsoluexitchaintbis}
\end{eqnarray}
and
\begin{eqnarray}
R_K(1,0) && =0 \nonumber \\
R_K(1,1) && =1 \nonumber \\
R_K(1,k \geq 2) && =1+\sum_{m=1}^{k-1} \prod_{n=1}^{m} \frac{ p_-(n)}{ p_+(n)}
\nonumber \\
R_K(1,K) && = 1+\sum_{m=1}^{K-1} \prod_{n=1}^{m} \frac{ p_-(n)}{ p_+(n)}
 = 1+ \frac{p_-(1)}{p_+(1)} + \frac{p_-(1)p_-(2)}{p_+(1)p_+(2)} + ... + \frac{p_-(1) p_-(2)...p_-(K-1)}{p_+(1) p_+(2)...p_+(K-1)}
\label{rsoluexitchaint}
\end{eqnarray}

Plugging this solution into 
Eq \ref{eigenperturbprojexpliextremfinal} yields
\begin{eqnarray}
 \lambda_1  &&   
  \simeq  G _{K}  e^{- \beta J_K}  \left(1  -  \frac{R_K(1,K-1)}{R_K(1,K)}
- \frac{q(0)}{q(K) } \frac{1}{R_0(0,K-1)}   \right)
\nonumber \\
&&  \simeq  G _{K}  e^{- \beta J_K} \frac{1+e^{ 2 \beta J_a} }{R_0(0,K-1)}  
\label{lambda1final}
\end{eqnarray}
so  the renormalized amplitude of Eq. \ref{ggenetree} becomes
 \begin{eqnarray}
G_R &&  =\frac{ \lambda_1 }
{ 2 \cosh    \beta  J_a   } 
 =  G _{K}  e^{ \beta (J_a- J_K) } \frac{1 }{R_0(0,K-1)}  
\label{ggenetreebis}
\end{eqnarray}

So using Eq. \ref{psizeroKtreevcompob}
we need to compute the ratios of the two probabilities of Eq. \ref{ppm}
\begin{eqnarray}
\frac{ p_+(k) }{p_-(k)} =  \frac{ G_{k+1} 
  T_0^2(S_a,...,S_a,S_{k+1}=0,-S_a,...,-S_a)}
{G_{k}   T_0^2(S_a,...,S_a,S_k=0,-S_a,...,-S_a)}
=  \frac{ G_{k+1}   2 \cosh \beta (\sum_{i=1}^{k}  J_{i} - \sum_{i=k+2}^K  J_{i} +J_a ) }
{G_{k}  2 \cosh \beta (\sum_{i=1}^{k-1}  J_{i} - \sum_{i=k+1}^K  J_{i} +J_a )}
\label{ratioppm}
\end{eqnarray}
as well as the products
\begin{eqnarray}
\prod_{k=m}^{K-1} \frac{ p_+(k)}{ p_-(k)} && = 
\frac{G_{K}  T_0^2(S_a,S_a,...,S_a+,S_K=0)}{G_{m} 
  T_0^2(S_a,S_a,...,S_a,S_m=0,-S_a,-S_a,...,-S_a)}
= \frac{G_{K}  2 \cosh \beta (\sum_{i=1}^{K-1}  J_{i}  +J_a )  }{G_{m}  2 \cosh \beta (\sum_{i=1}^{m-1}  J_{i} - \sum_{i=m+1}^K  J_{i} +J_a ) }
\label{productsratioppm}
\end{eqnarray}

Our conclusion is thus that the dynamical path $(1,2,..,K)$
has a renormalized amplitude $ G_R^{(1,2,..,K)}$ given by
 \begin{eqnarray}
\frac{1}{G_R^{(1,2,..,K)}} &&  =  \frac{ e^{ \beta ( J_K-J_a ) }  }{ G _{K}   } R_0(0,K-1) 
 \nonumber \\
&& =  \frac{ e^{ \beta ( J_K-J_a ) }  }{ G _{K}   }
 \left[ 1+\sum_{m=1}^{K-1} \frac{G_{K}  2 \cosh \beta (\sum_{i=1}^{K-1}  J_{i}  +J_a )  }
{G_{m}  2 \cosh \beta (\sum_{i=1}^{m-1}  J_{i} - \sum_{i=m+1}^K  J_{i} +J_a ) } \right]
 \nonumber \\
&& \simeq  e^{ \beta ( J_K-J_a ) }  2 \cosh \beta (\sum_{i=1}^{K-1}  J_{i}  +J_a ) 
\sum_{m=1}^{K} \frac{1  }{G_{m}  2 \cosh \beta (\sum_{i=1}^{m-1}  J_{i} - \sum_{i=m+1}^K  J_{i} +J_a ) } 
 \nonumber \\
&& \simeq e^{ \beta  \sum_{i=1}^{K}  J_{i}  } 
\sum_{m=1}^{K} \frac{1  }{G_{m} e^{ \beta \vert \sum_{i=1}^{m-1}  J_{i} - \sum_{i=m+1}^K  J_{i} +J_a \vert } } 
\label{ggenetreefin}
\end{eqnarray}

In terms of the dynamical barriers near zero temperature introduced in Eq. \ref{bi}
\begin{eqnarray}
 G_i  \oppropto_{ \beta \to +\infty} e^{ - \beta {\cal B}_i }
\label{bibis}
\end{eqnarray}
Eq. \ref{ggenetreefin} yields that
the renormalized barrier ${\cal B}_R^{(1,2,..,K)}$ 
associated to the dynamical path $(1,2,..,K)$ near zero temperature
is given by
 \begin{eqnarray}
{\cal B}_R^{(1,2,..,K)} && \equiv \lim_{\beta \to +\infty} \frac{  \ln \frac{1}{G_R^{(1,2,..,K)}}}{\beta} 
 = \max_{ 1 \leq m \leq K} \left[ {\cal B}_m + \sum_{i=1}^{K}  J_{i} - \left\vert \sum_{i=1}^{m-1}  J_{i} - \sum_{i=m+1}^K  J_{i} +J_a \right\vert \right]
\nonumber \\ && 
= \max_{ 1 \leq m \leq K} \left[ {\cal B}_m + J_m -J_a + 2 \min \left(\sum_{i=1}^{m-1}  J_{i}+J_a ; 
\sum_{i=m+1}^K  J_{i}  \right) \right]
\label{brpath}
\end{eqnarray}

\subsection{ Optimization over the $K!$ dynamical paths }

We now have to consider the $K!$ possible dynamical paths :
for a given permutation $\pi$ of the $K$ renormalized spins, 
the dynamical barrier ${\cal B}_R^{(\pi(1),\pi(2),..,\pi(K))}$
 associated to the path $(\pi(1),\pi(2),..,\pi(K))$
reads by adapting Eq \ref{brpath}
 \begin{eqnarray}
{\cal B}_R^{(\pi(1),\pi(2),..,\pi(K))} 
 = 
\max_{ 1 \leq m \leq K} \left[ {\cal B}_{\pi(m)} + J_{\pi(m)} -J_a 
+ 2 \min \left(\sum_{i=1}^{m-1}  J_{\pi(i)}+J_a ; 
\sum_{i=m+1}^K  J_{\pi(i)}  \right) \right]
\label{brpathpi}
\end{eqnarray}

We now have to choose the dynamical path, i.e. the
permutation $\pi$ leading to the smallest barrier. So
the final renormalization rule is that the renormalized barrier ${\cal B}_R $
is given by the minimum of Eq. \ref{brpathpi} 
over the $K!$ possible permutations
 \begin{eqnarray}
{\cal B}_R && \equiv \min_{\pi} \left( {\cal B}_R^{(\pi(1),\pi(2),..,\pi(K))} \right)
\nonumber \\ &&  = 
\min_{\pi} \left( \max_{ 1 \leq m \leq K} \left[ {\cal B}_{\pi(m)} + J_{\pi(m)} -J_a 
+ 2 \min \left(\sum_{i=1}^{m-1}  J_{\pi(i)}+J_a ; 
\sum_{i=m+1}^K  J_{\pi(i)}  \right) \right] \right)
\label{brtot}
\end{eqnarray}

\subsection{ Limit of the pure case} 

\label{sec_pure}

In the pure case, all couplings $J_i$ have the same value $J_0$
and all renormalized boundary spins $S_i$ have also the same barrier ${\cal B}_i={\cal B} $,
so that the renormalization rule of Eq. \ref{brtot}
reduces to
 \begin{eqnarray}
{\cal B}_R &&   = {\cal B}  
+ 2 J_0
 \max_{ 1 \leq m \leq K} \left[  \min \left( m  ; K-m \right) \right] 
\label{brtotpurdef}
\end{eqnarray}
For even $K$, the maximum is reached for $m=K/2$,
whereas for odd $K$, the maximum is reached for $m=(K \pm 1)/2$,
so that Eq. \ref{brtotpurdef} yields
 \begin{eqnarray}
{\cal B}_R  && =  {\cal B}+K J_0 \ \  \ \ \ \ \ \  \ \ \ \ {\rm for \ even \ } K
 \nonumber \\
 && = {\cal B}+(K-1) J_0 \ \ \ \ \ \ {\rm for \ odd \ } K
\label{brtotpur}
\end{eqnarray}
This recurrence thus leads to the following linear growth
 \begin{eqnarray}
{\cal B}_{n} - {\cal B}_{n=0} && =  n K J_0 \ \  \ \  \ \ \ \  \ \ \ \ {\rm for \ even \ } K
 \nonumber \\
 && =  n (K-1) J_0 \ \ \ \ \ \ {\rm for \ odd \ } K
\label{barriertree}
\end{eqnarray}
with the number $n$ of generations, in agreement with the more detailed results of \cite{us_rgdyn} containing explicit combinatorial prefactors coming from the degeneracy of the $K!$
dynamical paths. 
The linear growth with the number $n$ of generations is
 in agreement with previous works of physicists \cite{henley,melin,montanari}
and of mathematicians \cite{leng,yanna,marti,berger}.
It turns out that the slope $(K-1) J_0 $ for odd $K$ of Eq. \ref{barriertree}
coincides with the slope obtained in \cite{melin}, 
where a so-called 'disjoint strategy' is optimal,
whereas the slope $K J_0 $ for even $K$ of Eq. \ref{barriertree} 
differs from the slope $J(K-1)$ obtained in \cite{henley,melin},
where a so-called 'non-disjoint strategy' is optimal. 
We refer to Refs \cite{henley,melin,montanari,leng,yanna} for more explanations
on the differences between disjoint/non-disjoint strategies. 
Here it is clear that the renormalization procedure
making coherent clusters of spins within sub-trees corresponds to the disjoint strategy.
Within the disjoint strategy, the renormalization of Eq. \ref{brtot} for dynamical barriers
is close to the recursions written in Refs \cite{henley,melin,montanari,leng,yanna}, even if
there exists a difference : in our approach based on the quantum perturbation theory of section 
\ref{sec_reminder}, the spin $S$ is considered to be at equilibrium with respect to its neighbors,
whereas in previous approaches \cite{henley,melin,montanari,leng,yanna}, the question of the time
where the roots flips with respect to the flips of the sub-trees has been taken into account differently
(see the notions of 'directed barriers', of 'anchored barriers', etc...).
Let us also mention that the case of 'disorder in the degrees of the nodes' has been studied by Henley \cite{henley}, via the model of critical percolation on the regular Cayley tree.

\section { RG flow for the probability distribution of dynamical barriers  } 

\label{sec_rgflow}

\subsection{ Recurrence for the joint distribution of dynamical barriers and couplings} 

The renormalization rule of Eq. \ref{brtot} for dynamical barriers
yields that the joint probability distribution ${\cal P}_n({\cal B},J)$ of the barrier $B$ and the coupling
$J$ evolves according to the following iteration in terms of the probability distribution $\rho(J)$
of the random ferromagnetic couplings
 \begin{eqnarray}
{\cal P}_{n+1}({\cal B}_R , J_a)  = && \rho(J_a) \left( \prod_{i=1}^K \int d{\cal B}_i dJ_i {\cal P}_n({\cal B}_i ,J_i) \right) 
 \nonumber \\ 
&& \delta \left[{\cal B}_R + J_a -  \min_{\pi} \left( \max_{ 1 \leq m \leq K} \left[ {\cal B}_{\pi(m)} + J_{\pi(m)} 
+ 2 \min \left(\sum_{i=1}^{m-1}  J_{\pi(i)}+J_a ; 
\sum_{i=m+1}^K  J_{\pi(i)}  \right) \right] \right)  \right]
\label{itercalpn}
\end{eqnarray}
with the partial normalization
 \begin{eqnarray}
\int d{\cal B} P_{n}({\cal B}, J)  =  \rho(J) 
\label{normacal}
\end{eqnarray}

The initial condition for the Glauber dynamics of Eq. \ref{glauber}
near zero temperature  $G^{ini}_{Glauber} [h] = \frac{1}
{ 2 \cosh \left( \beta  h \right) } \simeq e^{- \beta \vert h \vert } $
corresponds for an initial boundary spin $S_i$ of the tree submitted to the 
local field $h=J_i S$
to the dynamical barrier ${\cal B}_i = J_i$, so that the joint distribution at generation
 zero reads
 \begin{eqnarray}
{\cal P}^{glauber}_{n=0}({\cal B} , J) &&  = \rho(J) \delta({\cal B}-J)
\label{iniglaubercal}
\end{eqnarray}

The form of Eq. \ref{itercalpn} actually shows that it is more convenient to replace
the variable ${\cal B}$ by $B={\cal B}+J $, so that the joint distribution $P_n(B,J)$
evolves according to
 \begin{eqnarray}
P_{n+1}(B , J_a)  = && \rho(J_a) \left( \prod_{i=1}^K \int dB_i dJ_i P_n(B_i ,J_i) \right) 
 \nonumber \\ 
&& \delta \left[B -  \min_{\pi} \left( \max_{ 1 \leq m \leq K} \left[ B_{\pi(m)}  
+ 2 \min \left(\sum_{i=1}^{m-1}  J_{\pi(i)}+J_a ; 
\sum_{i=m+1}^K  J_{\pi(i)}  \right) \right] \right)  \right]
\label{iterpn}
\end{eqnarray}
with the same partial normalization as in Eq. \ref{normacal}
 \begin{eqnarray}
\int d B P_{n}(B , J)  =  \rho(J) 
\label{norma}
\end{eqnarray}
and the initial condition (Eq. \ref{iniglaubercal})
 \begin{eqnarray}
P^{glauber}_{n=0}(B , J) &&  = \rho(J) \delta(B-2 J)
\label{iniglauber}
\end{eqnarray}

\subsection{ Recurrence for integrated probability distributions } 

To factorize the minimum and maximum functions involved in the delta function of Eq. \ref{iterpn}, it is convenient to introduce the two complementary integrated distributions
 \begin{eqnarray}
F_{n}( B , J) && \equiv  \int_{ B}^{+\infty} dB' P_{n}(B'  , J) 
 \nonumber \\ 
G_{n}( B , J) && \equiv  \int_{-\infty}^{ B} d B' P_{n}( B' , J) =  \rho(J) -  F_n( B,J)
\label{defgn}
\end{eqnarray}
as well as the notation
 \begin{eqnarray}
S^{(\pi)}_{\pi(m)} \equiv  2 \min \left(\sum_{i=1}^{m-1}  J_{\pi(i)}+J_a ; 
\sum_{i=m+1}^K  J_{\pi(i)}  \right) 
\label{Smpi}
\end{eqnarray}
Then the recurrence of Eq. \ref{iterpn} yields
 \begin{eqnarray}
F_{n+1}(B , J_a) &&  \equiv \int_{ B}^{+\infty} d B' P_{n+1}(B' , J_a)   
 \nonumber \\ && = \rho(J_a) \left( \prod_{i=1}^K \int dB_i dJ_i P_n(B_i ,J_i) \right)  
\theta \left[ B \leq \min_{\pi} \left( \max_{ 1 \leq m \leq K} \left[ {\cal B}_{\pi(m)}
 + S^{(\pi)}_{\pi(m)}  \right] \right) \right]
 \nonumber \\ && = \rho(J_a) \left( \prod_{i=1}^K \int dB_i dJ_i P_n(B_i ,J_i) \right) \
\prod_{\pi} \theta \left( B \leq  \max_{ 1 \leq m \leq K}
 \left[ B_{\pi(m)} +S^{(\pi)}_{\pi(m)}     \right] \right)
\label{iterfn}
\end{eqnarray}

Let us consider that we have ordered the $K!$ permutations $\pi$,
so that the last product of Eq. \ref{iterfn} can be expanded as 
 \begin{eqnarray}
  \prod_{\pi} \theta \left[ B \leq  \max_{ 1 \leq m \leq K} \left[ B_{\pi(m)} + S^{(\pi)}_{\pi(m)}  \right]   \right]
&&
  =  \prod_{\pi} \left( 1-  \theta \left[ B \geq  \max_{ 1 \leq m \leq K} 
\left[ B_{\pi(m)} + S^{(\pi)}_{\pi(m)}  \right]   \right] \right)
 \nonumber \\ && 
 = 1+\sum_{p=1}^{K!} (-1)^p \sum_{\{\pi_1<\pi_2<..<\pi_p\}} \prod_{q=1}^p
  \theta \left[ B \geq  \max_{ 1 \leq m \leq K} \left[ B_{\pi_q(m)}
 + S^{(\pi_q)}_{\pi_q(m)}   \right] \right]
 \nonumber \\ && 
 = 1+\sum_{p=1}^{K!} (-1)^p \sum_{\{\pi_1<\pi_2..<\pi_p\}} \prod_{q=1}^p
\prod_{m=1}^K   \theta \left[ B \geq   B_{\pi_q(m)}
 + S^{(\pi_q)}_{\pi_q(m)}   \right]
\label{prodpi}
\end{eqnarray}
For each permutation $\pi_q$, we may replace the product over $m=1,2,..,K$,
by the product over $k=\pi_q(m)$ to obtain
 \begin{eqnarray}
  \prod_{\pi} \theta \left[ B \leq  \max_{ 1 \leq m \leq K} \left[ B_{\pi(m)} + S^{(\pi)}_{\pi(m)}
 \right]   \right]
&&  = 1+ \sum_{p=1}^{K!} (-1)^p \sum_{\{\pi_1<\pi_2..<\pi_p\}} \prod_{q=1}^p
\prod_{k=1}^K   \theta \left[ B \geq   B_{k}
 + S_{k}^{(\pi_q)}  \right]
 \nonumber \\ && 
 = 1+\sum_{p=1}^{K!} (-1)^p \sum_{\{\pi_1<\pi_2..<\pi_p\}} \prod_{k=1}^K \prod_{q=1}^p
  \theta \left[ B \geq   B_{k} + S_{k}^{(\pi_q)}  \right]
 \nonumber \\ && 
 = 1+\sum_{p=1}^{K!} (-1)^p \sum_{\{\pi_1<\pi_2<..<\pi_p\}} \prod_{k=1}^K 
  \theta \left[ B \geq   B_{k} + \max_{ 1 \leq q \leq p } S_{k}^{(\pi_q)}
  \right]
\label{prodpibis}
\end{eqnarray}

Then Eq \ref{iterfn} becomes
 \begin{eqnarray}
F_{n+1}(B , J_a) && = \rho(J_a) \left( \prod_{i=1}^K \int dB_i dJ_i P_n(B_i ,J_i) \right)
\left[ 1+ \sum_{p=1}^{K!} (-1)^p \sum_{\{\pi_1<\pi_2..<\pi_p\}} \prod_{k=1}^K 
  \theta \left[ B \geq   B_{k} + \max_{ 1 \leq q \leq p }  S_{k}^{(\pi_q)} \right]\right]
 \nonumber \\ && 
=\rho(J_a)+ \rho(J_a)  \sum_{p=1}^{K!} (-1)^p \sum_{\{\pi_1<\pi_2<..<\pi_p\}} 
 \left( \prod_{k=1}^K \int dB_k dJ_k P_n(B_k ,J_k) 
\theta \left[   B_{k} \leq B -  \max_{ 1 \leq q \leq p } S_{k}^{(\pi_q)}  \right]\right)
 \nonumber \\ && 
=\rho(J_a)+ \rho(J_a)  \sum_{p=1}^{K!} (-1)^p \sum_{\{\pi_1<\pi_2<..<\pi_p\}} 
 \left( \prod_{k=1}^K  dJ_k G_n(B -  \max_{ 1 \leq q \leq p } S_{k}^{(\pi_q)} ,J_k) \right)
\label{iterfnclosed}
\end{eqnarray}
So we obtain the following closed recurrence for
the integrated probability $G_n$ of Eq. \ref{defgn}
 \begin{eqnarray}
G_{n+1}(B , J_a) && 
= - \rho(J_a)  \sum_{p=1}^{K!} (-1)^p \sum_{\{\pi_1<\pi_2..<\pi_p\}} 
 \left( \prod_{k=1}^K \int  dJ_k 
G_n(B -  \max_{ 1 \leq q \leq p } S_{k}^{(\pi_q)} ,J_k)   \right)
\label{itergnclosed}
\end{eqnarray}
with the initial condition of Eq. \ref{iniglauber}
 \begin{eqnarray}
G^{glauber}_{n=0}(B , J)  = \int_{-\infty}^B db P^{glauber}_{n=0}(b , J)  = \rho(J) \theta(B-2J)
&& =  0 \ \ {\rm for } \ \ B<2 J 
 \nonumber \\
 && = \rho(J) \ \ {\rm for } \ \ B > 2 J
\label{iniglauberbis}
\end{eqnarray}

\subsection{ Traveling-wave solution } 

\label{sec_tw}

For disordered models defined on Cayley trees , it is very common to find that the probability
distribution $P_n(A)$ of some observable $A$ propagates with a speed $v$ and
a fixed shape $P^*$
as the number $n$ of generations grows
\begin{eqnarray}
P_n(A) \simeq P^* (A-nv)
\label{twsol}
\end{eqnarray}
 This means that the average value grows linearly with $n$,
whereas the width around this averaged value remains finite.
This property was discovered by Derrida and Spohn \cite{Der_Spohn}
on the specific example of the directed polymer in a random medium,
where the observable $A$ of interest is the free-energy,
and was then found in various other statistical physics models
defined on Cayley trees  \cite{majumdar}.
This traveling-wave propagation of probability distributions
have also been found in quantum models defined on Cayley trees,
in particular in the Anderson localization problem
\cite{abou,zirnbauer,efetov,bell,us_cayley}.
The conclusion is thus that the recursion relations that can be written
for observables of disordered models defined on trees naturally lead to
the traveling wave propagation of the corresponding probability distributions.

So here, it is natural to expect that the solution of Eq. \ref{itergnclosed} starting from the initial condition
of Eq. \ref{iniglauberbis} will be a traveling-wave with some velocity $v$
 \begin{eqnarray}
G_n(B,J) \simeq G^*(b \equiv B- n v ,J)
\label{twpn}
\end{eqnarray}
where the stable shape
$G^*(b,J)$ of the front satisfies the equation
 \begin{eqnarray}
G^*(b-v , J_a) && 
= - \rho(J_a)  \sum_{p=1}^{K!} (-1)^p \sum_{\{\pi_1<\pi_2..<\pi_p\}} 
 \left( \prod_{k=1}^K \int  dJ_k 
G^*(b -  \max_{ 1 \leq q \leq p } S_{k}^{(\pi_q)} ,J_k)   \right)
\label{fixedgnclosed}
\end{eqnarray}
and the following boundary conditions at infinity
 \begin{eqnarray}
 G^*(b  , J) && \opsimeq_{b \to -\infty} 0
\nonumber \\ 
 G^*(b  , J) && \opsimeq_{b \to +\infty} \rho(J)
\label{glimits}
\end{eqnarray}

This means that a barrier $B_n$ for a sub-tree of $n$ generations reads
 \begin{eqnarray}
B_n= n v + b
\label{tw}
\end{eqnarray}
where $b$ is a random variable of order $O(1)$.
In the following sections, we have checked numerically for $K=2$ and $K=3$
that the statistics of dynamical barriers indeed follow the traveling-wave form
of Eq. \ref{twpn}.

\section{ CASE OF BRANCHING NUMBER $K=2$ }

\label{sec_k2}

\subsection{ Renormalization rule for dynamical barriers } 

Here there are only $K!=2$ permutations $\pi_1=(1,2)$ and $\pi_2=(2,1)$, so 
the barriers associated to these two permutations are (Eq \ref{brpathpi})
 \begin{eqnarray}
{\cal B}_R^{(1,2)} 
&& = \max \left[ {\cal B}_{1} + J_1-J_a+2 \min (J_2,J_a) ; {\cal B}_{2} + J_2  - J_a   \right]
 \nonumber \\
{\cal B}_R^{(2,1)} 
&& = \max \left[ {\cal B}_{2} + J_2-J_a+2 \min (J_1,J_a) ; {\cal B}_{1} + J_1  - J_a   \right]
\label{brpathpiK2}
\end{eqnarray}
and the optimization of Eq. \ref{brtot} reads
 \begin{eqnarray}
{\cal B}_R +J_a \equiv \min \left(  \max \left[ {\cal B}_{1} + J_1+2 \min (J_2,J_a) ; {\cal B}_{2} + J_2    \right] ; \max \left[ {\cal B}_{2} + J_2+2 \min (J_1,J_a) ; {\cal B}_{1} + J_1    \right] \right)
\label{brtotK2}
\end{eqnarray}

\subsection{ Recurrence for the integrated probability }

 For $K=2$, we have to consider the two permutations $\pi_1=(1,2)$ and $\pi_2=(2,1)$, 
so the variables of Eq.  \ref{Smpi} reads
 \begin{eqnarray}
S_{1}^{(\pi_1)} &&  = 2 \min(J_2,J_a)
 \nonumber \\
S_{2}^{(\pi_1)}  && = 0
 \nonumber \\
S_{1}^{(\pi_2)}  && = 0
 \nonumber \\
S_{2}^{(\pi_2)}  && = 2 \min(J_1,J_a)
\label{SmpiK2}
\end{eqnarray}
and Eq \ref{itergnclosed} becomes
 \begin{eqnarray}
 G_{n+1}(B , J_a)  
&& =  \rho(J_a)  
  \int  dJ_1 \int  dJ_2 
G_n(B -   S_{1}^{(\pi_1)}  ,J_1) 
G_n(B -   S_{2}^{(\pi_1)}  ,J_2) 
 \nonumber \\ && 
+ \rho(J_a)  
 \int  dJ_1  \int  dJ_2 
G_n(B -   S_{1}^{(\pi_2)}  ,J_1)   
G_n(B -   S_{2}^{(\pi_2)} ,J_2)   
 \nonumber \\ &&
 - \rho(J_a) 
  \int  dJ_1    \int  dJ_2 
G_n(B -  \max ( S_{1}^{(\pi_1)}
 , S_{1}^{(\pi_2)}) ,J_1)  
G_n(B -  \max ( S_{2}^{(\pi_1)}
 , S_{2}^{(\pi_2)})  ,J_2) 
 \nonumber \\
&& =  \rho(J_a)  
  \int  dJ_1 \int  dJ_2 
G_n(B -  2 \min(J_2,J_a)     ,J_1) 
G_n(B   ,J_2) 
 \nonumber \\ && 
+ \rho(J_a)  
 \int  dJ_1  \int  dJ_2 
G_n(B    ,J_1)   
G_n(B - 2 \min(J_1,J_a)      ,J_2)   
 \nonumber \\ &&
 - \rho(J_a) 
  \int  dJ_1    \int  dJ_2 
G_n(B -  2 \min(J_2,J_a) ,J_1)  
G_n(B -  2 \min(J_1,J_a)   ,J_2) 
\label{itergnclosedK2}
\end{eqnarray}
i.e. finally, the integrated probability $G_n$ satisfies the recurrence
 \begin{eqnarray}
 G_{n+1}(B , J_a) 
&& =  \rho(J_a)  
  \int  dJ_1 \int  dJ_2 
G_n(B -2 \min(J_2,J_a)  )   ,J_1) 
\left[ 2 G_n(B  ,J_2) - G_n(B -2  \min(J_1,J_a)   ,J_2)  \right]
\label{itergnclosedK2final}
\end{eqnarray}

\subsection{ Traveling-wave Ansatz }

The traveling-wave Ansatz of Eq. \ref{twpn} yields the following equation
for  the stable shape
$G^*(b,J)$ of the front 
 \begin{eqnarray}
 G^*(b -v , J_a) 
&& =  \rho(J_a)  
  \int  dJ_1 \int  dJ_2 
G^*(b -2 \min(J_2,J_a)  )   ,J_1) 
\left[ 2 G^*(b  ,J_2) - G^*(b -2  \min(J_1,J_a)   ,J_2)  \right]
\label{fixedgnclosedK2}
\end{eqnarray}
with the boundary conditions of Eq. \ref{glimits}

\subsection{ Weak-disorder expansion  } 

For the pure case $\rho(J)=\delta(J-J_0)$ discussed in section \ref{sec_pure},
the solution is simply 
 \begin{eqnarray}
 G^*_{pure}(b , J) && =  \delta(J-J_0) \theta(b) = \rho(J) \theta(b)
\label{frontpureK2}
\end{eqnarray}
with the velocity of Eqs \ref{brtotpur} and \ref{barriertree}
 \begin{eqnarray}
v_{pure}= 2 J_0
\label{vpure}
\end{eqnarray}

Let us now consider the case of weak disorder, where the disorder distribution
$\rho(J)$ displays a small width around $J_0$. 
Then we expect that the width of the barrier distribution will also be small
with respect to $(2 J_0)$. In the region where $G^*(b -v , J_a)  $ is small,
 the appropriate linearization of Eq. \ref{fixedgnclosedK2}
consists in replacing $ G^*(b  ,J_2)$ by its asymptotic behavior $ \rho(J_2)$ (Eq \ref{glimits})
and by neglecting $G^*(b -2  \min(J_1,J_a)   ,J_2) $ that would give a quadratic contribution, so that we obtain in this weak-disorder regime
 \begin{eqnarray}
 G^*(b -v , J_a) && \simeq 2  \rho(J_a)  
  \int  dJ_1 \int  dJ_2 \rho(J_2)
G^*(b -2 \min(J_2,J_a)     ,J_1) 
\label{fixedgnclosedK2linear}
\end{eqnarray}
This means that the integral
 \begin{eqnarray}
 g^*(b ) && \equiv \int dJ G^*(b , J)
\label{defgstar}
\end{eqnarray}
satisfies the closed linear equation
 \begin{eqnarray}
 g^*(b -v) && \simeq 2 \int dJ_a  \rho(J_a)  
   \int  dJ_2 \rho(J_2)
g^*(b -2 \min(J_2,J_a)  ) 
\label{closedK2linear}
\end{eqnarray}

The exponential shape of coefficient $\mu$
 \begin{eqnarray}
g^*(b) && \opsimeq_{b \to -\infty}  e^{\mu b}
\label{tailmu}
\end{eqnarray}
is then a solution of the linearization of Eq. \ref{closedK2linear}
if the velocity $v(\mu)$ satisfies
 \begin{eqnarray}
 e^{- \mu v(\mu) }  && = 2 \int dJ_a \rho(J_a)  \int  dJ_2 \rho(J_2)
 e^{- 2 \mu  \min ( J_2, J_a) } 
\label{Av}
\end{eqnarray}
The velocity as a function of $\mu$
 \begin{eqnarray}
v(\mu)  && = - \frac{1}{\mu}
 \ln \left[  2 \int dJ_a \rho(J_a)  \int  dJ_2 \rho(J_2) e^{- 2 \mu  \min ( J_2, J_a) } \right]
\label{vmufinal}
\end{eqnarray}
presents the two limiting behaviors
 \begin{eqnarray}
 v(\mu) && \opsimeq_{\mu \to 0} - \frac{ \ln 2}{ \mu } \to -\infty
\nonumber \\  
v(\mu) &&  \opsimeq_{\mu \to +\infty} 2 J_{min} +  \frac{\ln \mu }{\mu} \to  2 J_{min}
\label{vmuexplim}
\end{eqnarray}
where $J_{min}$ is the minimal value of the disorder distribution $\rho(J)$.
Between these two limits, there exists a maximum value
 $v^*$ at $\mu^*$ such that $ v'(\mu)=0$, and this is this velocity $v^*$
that will be dynamically selected \cite{vanSaarloos,brunetreview}.
It can be computed for various disorder distributions $\rho(J)$.
Let us now consider two explicit cases.

\subsubsection{ Exponential distribution for the disorder }

For the exponential distribution
 \begin{eqnarray}
\rho(J)=\theta(J-J_0) \frac{1}{\Delta} e^{- \frac{J-J_0}{\Delta}}
\label{rhoexp}
\end{eqnarray}
Eq  \ref{vmufinal} reads
 \begin{eqnarray}
 v(\mu)  
&& =  2 J_0 + \frac{1}{\mu} \ln \frac{1+\mu \Delta }{2}   
\label{vmuexp}
\end{eqnarray}
We are looking for the point $\mu^*$ where this function is maximum 
 \begin{eqnarray}
0 && =  v'(\mu_*)  = - \frac{1}{\mu_*^2} \ln \left(  \frac{1+\mu_* \Delta }{2} \right)+
 \frac{ \Delta  }{\mu_* ( 1+\mu_*\Delta ) }
\label{vmuexpderi}
\end{eqnarray}
i.e. $\mu_* = x/\Delta $ where $x \simeq3.31107$ is the root of the numerical equation
 \begin{eqnarray}
 \ln \left(  \frac{1+x}{2} \right) =  \frac{x }{  1+x }
\label{xsol}
\end{eqnarray}
The corresponding selected velocity reads
 \begin{eqnarray}
v_*= v(\mu_*)  
&& =  2 J_0 +  \frac{1}{\mu_*} \ln \left(  \frac{1+\mu_* \Delta }{2} \right)
\nonumber \\  
&& =  2 J_0 + \frac{\Delta }{x} \ln \left(  \frac{1+x }{2} \right) 
\nonumber \\  
&& =  2 J_0 + \frac{\Delta }{1+x} 
\label{vmuexpstar}
\end{eqnarray}
In conclusion, we obtain that the velocity grows linearly for weak disorder $\Delta$
 \begin{eqnarray}
v^*= 2 J_0 +  0.23195  \Delta 
\label{vmuexpstarnume}
\end{eqnarray}

\subsubsection{ Box distribution for the disorder  }

For the box distribution of width $\Delta$
 \begin{eqnarray}
\rho(J)= \frac{1}{\Delta}\theta(J_0 \leq J \leq J_0+\Delta)
\label{rhobox}
\end{eqnarray}
 Eq \ref{vmufinal} becomes
 \begin{eqnarray}
v(\mu)  && = 2 J_0  - \frac{1}{\mu} \ln \left[  \frac{e^{- 2 \mu \Delta }+2 \mu \Delta  -1 }{\mu^2\Delta^2}   \right]
\label{vmufinalbox}
\end{eqnarray}
The corresponding selected velocity $v^*=v(\mu^*)$ (where $v'(\mu^*)=0 $) grows again linearly for weak disorder $\Delta$
 \begin{eqnarray}
v^*= 2 J_0 + \Delta v_*  \simeq 2+ 0.206826 \Delta
\label{vmustarnume}
\end{eqnarray}

\subsection{ Numerical results obtained via the pool method } 

\label{sec_pool}

\begin{figure}[htbp]
 \includegraphics[height=6cm]{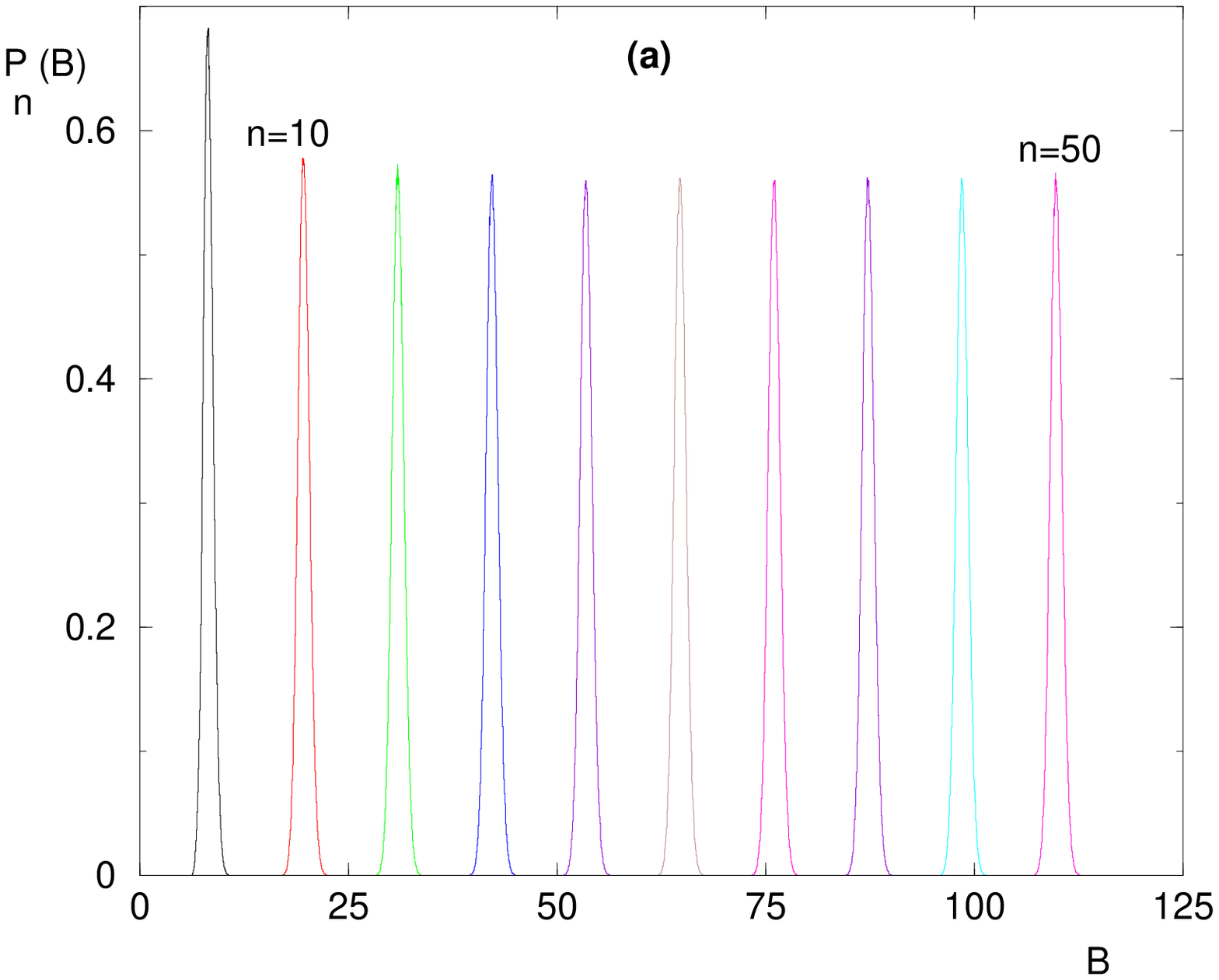}
\hspace{1cm}
 \includegraphics[height=6cm]{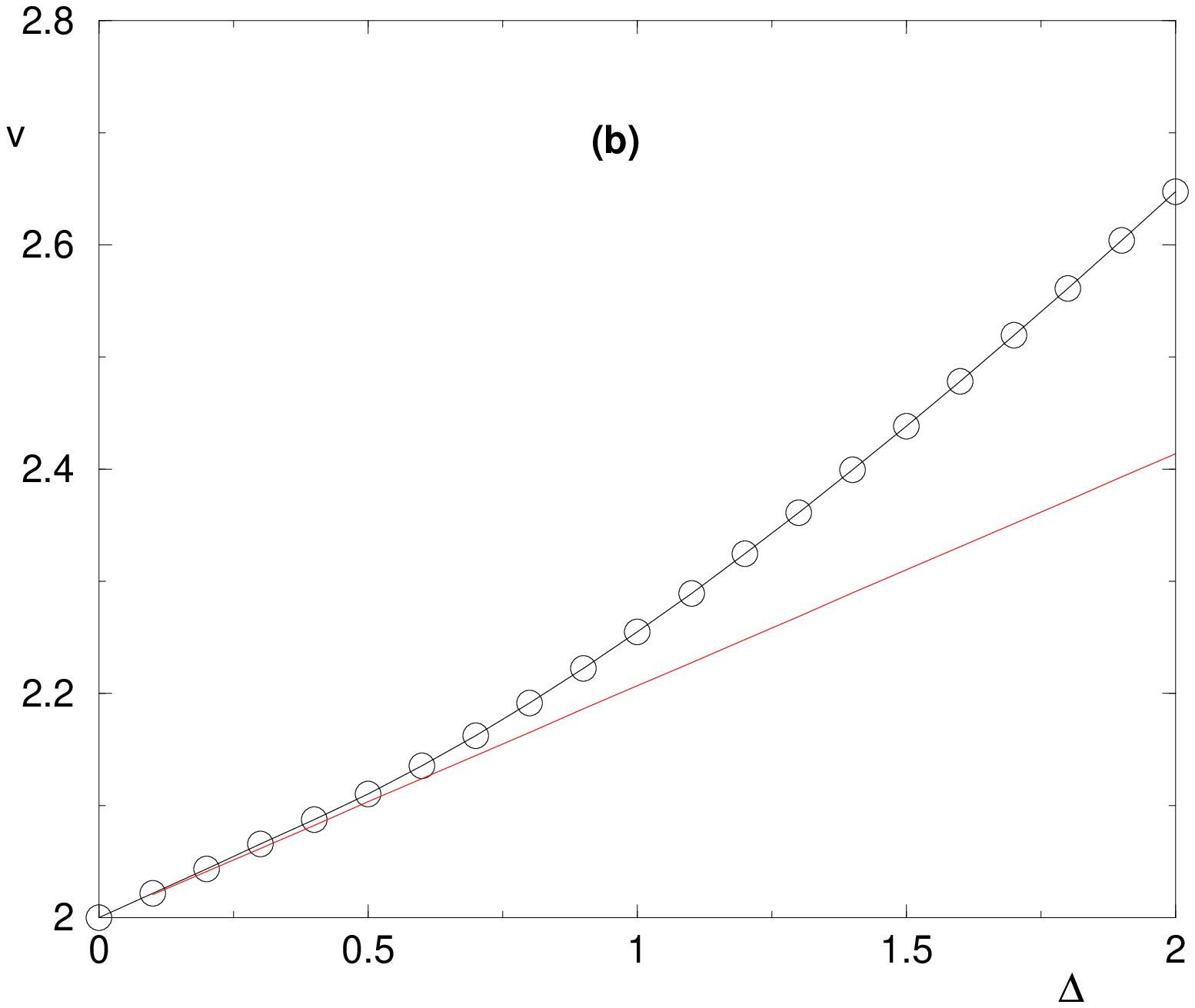}
\caption{ Cayley tree of branching ratio $K=2$ with the box distribution of disorder of Eq. \ref{rhobox} with $J_0=1$
(a) For $\Delta=1$ :  the probability distributions $P_n(B)$ of the dynamical barrier $B$
at generation $n=5,10,15,20,25,30,35,40,45,50$ correspond to a traveling-wave
with a fixed shape $P_n(B) \simeq P^*(b \equiv B- n v)$ (after an initial transient).
(b) Velocity $v$ of the traveling-wave as a function of the disorder width $\Delta$
 and agreement with the weak-disorder expansion at first order in $\Delta$
 of Eq. \ref{vmustarnume}. }
\label{figk2}
\end{figure}

If one wishes to study numerically real trees, one is limited to rather small
number $n$ of generations, because the number of sites grows exponentially in $n$.
From the point of view of convergence towards stable probability
distributions via recursion relations, it is thus better
 to use the so-called 'pool method'
that allows to study much larger number of generations.
The pool method has been used for disordered models on trees
\cite{abou,bell,us_cayley} or on hierarchical lattices
\cite{poolspinglass,Coo_Der,diamondpolymer}.
The idea of the pool method is the following :  
at each generation, one keeps the same number $M_{pool}$
of random variables to represent probability distributions.
Within our present framework, the probability distribution $P_n(B,J)$ at
 generation $n$ will be represented by a pool of $M_{pool}=10^6$ couples $(B_i,J_i)$.
To construct a new couple $(B_R,J_a)$ of generation $(n+1)$, one draws $K$ couples 
$(B_i,J_i)$ within the pool of generation $n$ and apply the rule of Eq. \ref{brtotK2}.

The numerical results obtained via the pool method for the box distribution
of the disorder of Eq. \ref{rhobox} with $J_0=1$
are shown on Fig. \ref{figk2}.
On Fig. \ref{figk2} (a), we display the convergence towards the traveling-wave form
for the finite disorder width $\Delta=1$.
On Fig. \ref{figk2} (b), we show that the velocity $v(\Delta)$ as a function of the disorder
strength has for tangent the weak-disorder expansion of Eq. \ref{vmustarnume} near $\Delta \to 0$.

\section {CASE OF BRANCHING NUMBER $K=3$  }

\label{sec_k3}

\subsection{ Renormalization rule for dynamical barriers } 

Here there are $K!=6$ permutations  so 
the barriers associated to these permutations are (Eq \ref{brpathpi})
 \begin{eqnarray}
{\cal B}_R^{\pi_1=(1,2,3)} 
&& 
 = 
\max \left[ 
{\cal B}_1 + J_1-J_a + 2 \min ( J_2+J_3;J_a ) ;
{\cal B}_2 + J_2-J_a +2 \min  (J_1 +J_a, J_3 );
{\cal B}_3 + J_3-J_a
\right]
 \nonumber \\
{\cal B}^{\pi_2=(2,1,3)}_R  && =
\max \left[ 
{\cal B}_2 + J_2-J_a + 2 \min ( J_1+J_3;J_a ) ;
{\cal B}_1 + J_1-J_a +2 \min  (J_2 +J_a, J_3 ) ;
{\cal B}_3 + J_3-J_a
\right]
 \nonumber \\
{\cal B}_R^{\pi_3=(1,3,2)} 
&& 
 = 
\max \left[ 
{\cal B}_1 + J_1-J_a + 2 \min ( J_2+J_3;J_a ) ;
{\cal B}_3 + J_3-J_a +2 \min  (J_1 +J_a, J_2 );
{\cal B}_2 + J_2-J_a
\right]
 \nonumber \\
{\cal B}_R^{\pi_4=(3,1,2)} 
&& 
 = 
\max \left[ 
{\cal B}_3 + J_3-J_a + 2 \min ( J_2+J_1;J_a ) ;
{\cal B}_1 + J_1-J_a +2 \min  (J_3 +J_a, J_2 ) ;
{\cal B}_2 + J_2-J_a
\right]
 \nonumber \\
{\cal B}^{\pi_5=(2,3,1)}_R  && =
\max \left[ 
{\cal B}_2 + J_2-J_a + 2 \min ( J_1+J_3;J_a ) ;
{\cal B}_3 + J_3-J_a +2 \min  (J_2 +J_a, J_1 ) ;
{\cal B}_1 + J_1-J_a
\right]
 \nonumber \\
{\cal B}^{\pi_6=(3,2,1)}_R  && =
\max \left[ 
{\cal B}_3 + J_3-J_a + 2 \min ( J_2+J_1;J_a ) ;
{\cal B}_2 + J_2-J_a +2 \min  (J_3 +J_a, J_1 ) ;
{\cal B}_1 + J_1-J_a
\right]
\label{brpathpiK3}
\end{eqnarray}
and the optimization of Eq. \ref{brtot} reads
 \begin{eqnarray}
{\cal B}_R \equiv \min \left( {\cal B}_R^{(1,2,3)} ; {\cal B}^{(2,1,3)}_R ; 
{\cal B}_R^{(1,3,2)} , {\cal B}_R^{(3,1,2)} , {\cal B}^{(2,3,1)}_R  , {\cal B}^{(3,2,1)}_R  \right)
\label{brtotK3}
\end{eqnarray}

The corresponding iteration of Eq \ref{itergnclosed} for the joint probability distribution is rather lengthy and not very illuminating, so we will not write it here,
but present instead our numerical results.

\subsection{ Numerical results obtained via the pool method } 

\begin{figure}[htbp]
 \includegraphics[height=6cm]{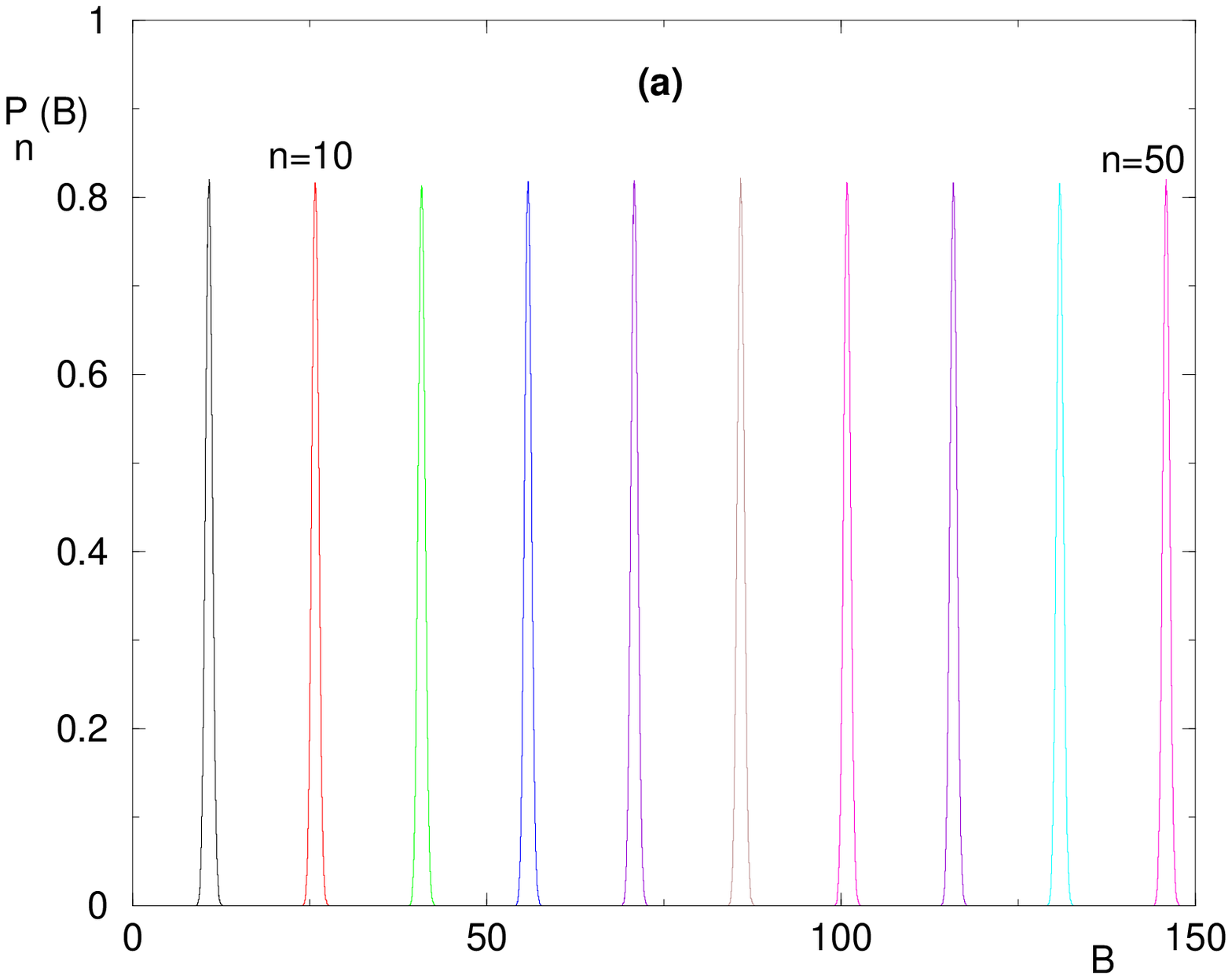}
\hspace{1cm}
 \includegraphics[height=6cm]{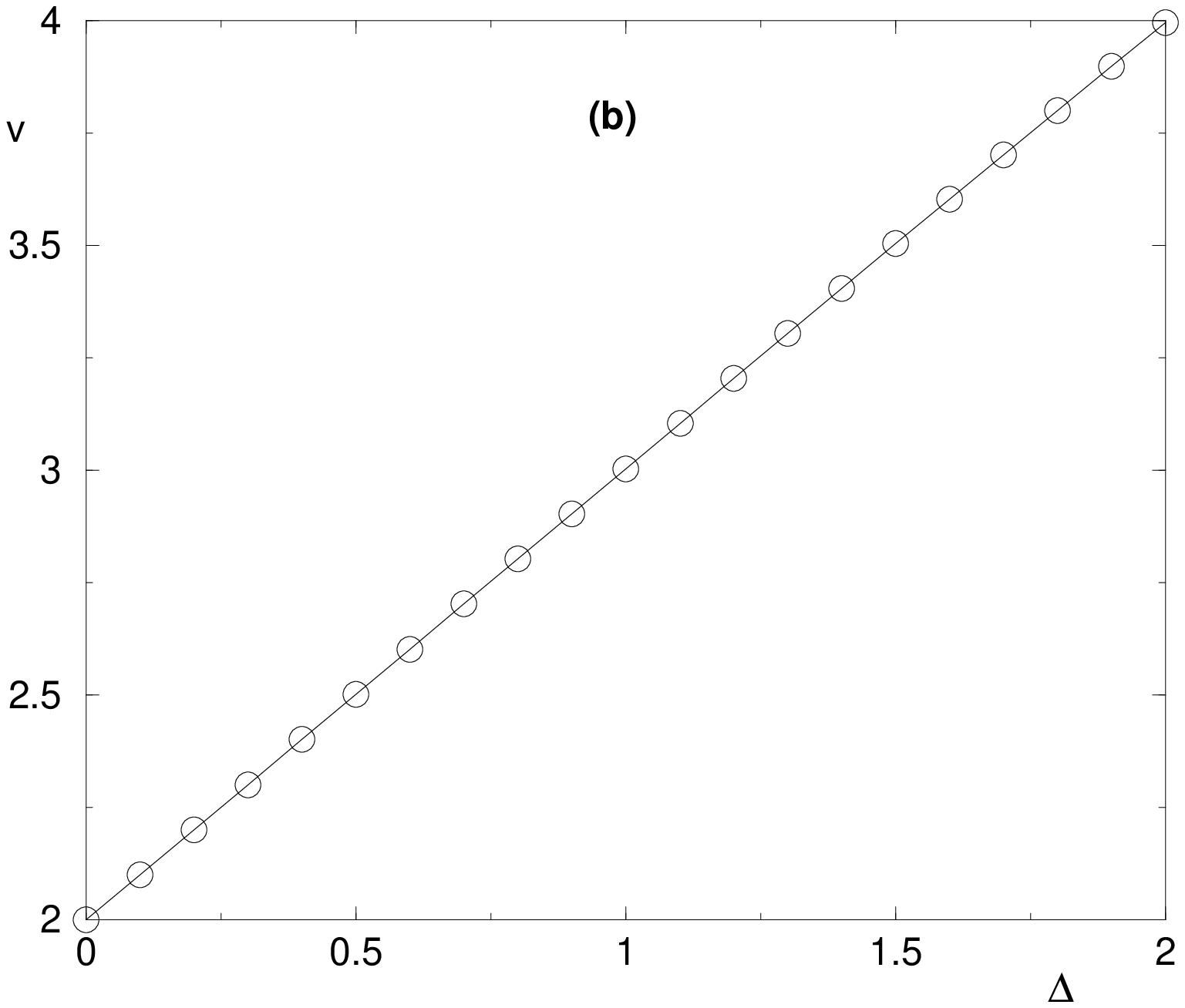}
\caption{ Cayley tree of branching ratio $K=3$ with the box distribution of disorder of Eq. \ref{rhobox} with $J_0=1$
(a) For $\Delta=1$ :  the probability distributions $P_n(B)$ of the dynamical barrier $B$
at generation $n=5,10,15,20,25,30,35,40,45,50$ correspond to a traveling-wave
with a fixed shape $P_n(B) \simeq P^*(b \equiv B- n v)$ .
(b) Velocity $v$ of the traveling-wave as a function of the disorder width $\Delta$
 }
\label{figk3}
\end{figure}

The numerical results obtained via the pool method for the box distribution
of the disorder of Eq. \ref{rhobox} with $J_0=1$
are shown on Fig. \ref{figk3}. We find again that the statistics of dynamical barriers is a traveling wave (see Fig. \ref{figk3} (a)). The corresponding velocity $v(\Delta)$ 
as a function of the disorder strength $\Delta$ is shown on Fig. 
\ref{figk3} (b).

\section {Conclusion }

\label{sec_conclusion}

To study the stochastic dynamics near zero-temperature of the random ferromagnetic Ising model on a Cayley tree of branching ratio $K$, we have applied the Boundary Real Space Renormalization procedure introduced in our previous work \cite{us_rgdyn} to derive the renormalization rule for dynamical barriers. The main outcome is that the probability distribution $P_n(B)$ of dynamical barrier for a subtree of $n$ generations converges for large $n$ towards some traveling-wave $P_n(B) \simeq P^*(B-nv) $, i.e. the width of the probability distribution remains finite around an average-value that grows linearly with the number $n$ of generations. We have presented numerical results for the branching ratios $K=2$ and $K=3$, and we have computed the weak-disorder expansion of the velocity $v$ for $K=2$.

As explained in section \ref{sec_tw}, the recursion relations that can be written
for observables of disordered models defined on trees naturally lead to
the traveling wave propagation of probability distributions.
So we expect that for other disordered statistical models defined on trees,
the statistics of dynamical barriers should be also described by travelling-waves.

\end{document}